# Big Science and Science Education: Steps towards an Authentic Partnership

**Stephen M. Pompea**[1]
NSF's NOIRLab, Tucson, Arizona, USA and Leiden University, Leiden, the Netherlands

**Pedro Russo**
Department of Science Communication & Society and Leiden Observatory
Leiden University, Leiden, the Netherlands

Address for correspondence: stephen.pompea@noirlab.edu

**Abstract**
Big science projects and facilities can move towards a less self-centered frame of reference as they strive to better identify and serve educational audiences. By doing this, their science education efforts will be more productive in general, and their service to local schools will be more effective. By developing an enlarged awareness of local educational needs, they will become better stewards and partners in their roles in the science education system. They will also become more valued and trustworthy neighbours to their local and cultural communities. We propose a practical way for large science organisations to organise their budgets and their allocation of staff time to greatly increase the effectiveness of their organisation in its contribution to local science education.

**Introduction**
Big science transforms our understanding of the natural world. It brings new data, knowledge, and wonder not only to scientists but to the broader science-interested public. It mobilizes entire sectors of society intellectually as well as commercially as it develops new projects and the new facilities that support them. The scientific results are conspicuous and consistent successes: Earthscope, the National Ignition Facility, the Large Hadron Collider, and the Human Genome Project have all transformed their branches

---
[1] Adjunct Faculty, James C. Wyant College of Optical Sciences and Department of Astronomy & Steward Observatory, University of Arizona, Tucson, Arizona, USA

of science. The largest projects involve tens of thousands of scientists from over a 100 countries. These projects, and the tools they have created for international collaboration, have transformed modern science by rapidly making enormous amounts of data available to large groups of scientists worldwide. The era of Big Science has transformed our relationship to science data. The title of one data-based project provides a very apt description of the era of Big Science: "From Petabytes to Megafolks in Milliseconds" (NSF 2022).

These big science initiatives are very familiar to the authors; we have communications and educational leadership roles in astronomical research institutes that are conducting large, ambitious projects such as the Event Horizon Telescope, the Dark Energy Survey, and the Legacy Survey of Space and Time. We have also been involved with the teams that are creating the next generation of highly capable and extremely large ground-based telescopes. These include optical-infrared telescopes with apertures of over 1500 square meters and radio telescope arrays whose collecting area is over a million square meters. We have also led projects in a different type of big science: enormous worldwide science education ventures involving over 100 countries. These projects include the International Year of Astronomy 2009, the IAU100 projects, and many other ambitious national and international science education initiatives (Russo & Christensen 2010; Pompea et al. 2010, van Dishoeck & Elmegreen 2018).

This article describes how big science projects and facilities can enlarge their outlook on science education and move towards a less self-centered frame of reference. By doing this, their science education efforts will be more focused and productive in serving local schools. They will also become better stewards and partners to the science education system, as well as more trustworthy neighbours to their local and cultural communities. We propose a way for large science organisations to organise their budgets and staff time to greatly increase their effectiveness in supporting local science education.

The first step to an enlarged perspective on what organisations can accomplish in science education is to explore the general area of science communication, community engagement, outreach, and science education. These sectors are often unified in research organisations as a functional department with a name that is a combination of these words. We believe that this combination or blurring of of duties is inadvisable, as it muddies the distinctions between the goals and results of the different efforts, often leading to a neglect of science education, and especially of local science education efforts. We also suggest that science organisations can benefit from adopting the strategy of making several specific commitments of money and staff time to educational efforts. Finally, we advise that big science organisations and their scientists begin to view science education efforts from a deeper ethical and cultural stewardship perspective, rather than from a more limited "business" model that tends to emphasize a quick return on educational efforts and investments.

We use the term "big science" to mean the lead and subsidiary science organisations and institutions that create and implement these projects. This includes the employees of these organisations: the scientists, engineers, project managers, and the leadership of these organisations. It also includes the funders and the supporting companies that make products used in doing science. In short, we are addressing everyone who works for these organisations that have an orientation towards enabling large science projects and their associated science advancements. This includes the individual scientists who work independently or at small organisations as well as university scientists who are part of a larger project. It also includes the support staff such as those who work in accounting, purchasing, safety, and human resources at these national labs, private labs, universities, and non-profits.

**Communications: A Success Story for Big Science**
Big science projects have advanced and transformed the field of science communication. They do a terrific job of communicating the excitement of the scientific process. They are highly successful at publicising new results to science stakeholders and to the public that play an important role in the support of their private or government funding. Science communicators are often among the most important first hires for a big science project under development. Science communication experts are essential players in helping outline and communicate a project's scientific objectives and in elucidating the exciting science questions that will be addressed. Science communication specialists are also needed for their professional skills during the lengthy and often convoluted odyssey of aligning stakeholders, participating institutions, foundations, and government agencies during the quest for funding and support.

The science of science communication has mature proven best practices which are essential elements in the creation of big science projects, organisations, and facilities (Christensen 2007; National Academy of Sciences 2017, 2018; Jamieson 2017;). Science communication professionals are highly creative in teaming with scientists to create exciting and successful wide-reaching projects designed to interest children and adults in science. For example, a worldwide project was hosted in many countries to allow the public to name a unique exoplanet and its host star. This IAU100 project was very successful in stimulating interest in exoplanets. with 113 national campaigns that involved nearly 780 thousand people (Mamajek et al. 2020).

**Science Education: A Story with Diffident Results**
If the science research and science communication efforts of big sciences are outstanding successes, their education efforts often fall short. There are many reasons for this. The initial educational efforts often have a significant activation energy and may start slowly or not at all. The international nature of these large projects can lead to a committee mentality where no one part of the project or organisation takes responsibility for the science education program. There is often a question of what could and should be done in education, with perhaps no one person on the initial team with solid answers or experience in this area. With distributed projects, there may be no mandate or marching orders towards doing science education, since the funding source and project scope are still under discussion. While science communication efforts are perceived as an essential need in a project's early phases, science education efforts are perceived to lack a compelling rationale and may be viewed as a bridge too far or a bridge to be crossed later.

The remote and rural nature of some of the large science facilities can lead to a sense of geographic disembodiment and a lack of identification with the local community, the region, or even the country in which the facility is placed. User facilities have an additional identity problem. They have many people coming and going and their operations managers may feel that they have enough difficulty in keeping the facility running continuously and in responding to the needs of the diverse users who cycle through the facility or its data. Thus, day-to-day operational concerns overwhelm a longer-term perspective.

The newest research facilities produce enormous amounts of data for a worldwide community of scientists who may never visit the physical location or understand the local situation. Related to this, in data intensive projects there often is a diminished relationship to the source of the data. The actual place that takes the data might be viewed more as a set of sensors or as a "peripheral" connected to the computers and its pipelines. There may be little concern for place or a sense of place as long as the data keeps flowing. Finally, with tight budgets, facilities are trying their best to focus on their core mission: doing science and providing science data. They work hard to eliminate mission creep, and to look for opportunities to divest themselves of any stranded tasks or requirements that no one seems to miss. They also try to minimize or ignore any new demands that don't arrive with additional money.

All of these factors influence the ability of an organisation to maintain an educational mission. If educational efforts are not well-defined and built in at the beginning of a project, they tend not to be treated seriously and they may be forsaken later.

These observations apply to the many players in the Big Science structure and not just to the national, federal or state research labs, user facilities, and university science departments and institutes. Other participants are the industrial contractors that make the steel and concrete, aerospace companies, engineering and architectural firms, and the support companies in information and data services. Many of these players have adopted a "not my job" approach to helping the science education ecosystem, even as they have solid communication, community engagement, and occasionally outreach efforts. The distinctions between these areas are important and will be discussed in detail later.

**Why The Education Mission Matters**
The lack of a solid, well-defined and measurable educational mission in many big science organisations has some serious long-term consequences for society and for the science enterprise. Even if the organisation is doing a great job of generating press releases, adding social media followers, making inspiring movies, enjoying the publicity from popular articles, and creating well-written science and funding justifications, they may not be perceived as a good neighbour or as a core member of their local community. Does this matter, though?

We believe that, in the longer term, it does matter, as there are very direct feedback loops that affect the organisation. It is often a surprise to science organisations when their trustworthiness and legitimacy to operate, expand, or to procreate is questioned by geographic or cultural communities nearby. This has happened in many places. In Hawai'i the building of a billion-dollar telescope on Maunakea (the TMT) has been challenged, delayed, and possibly stopped by the efforts of Hawai'ians with strong concerns about colonial attitudes by scientists and how the culture of science has allowed scientific advances to be made all too often at the expense of marginalized people (Kahanamoku et al. 2020; Prescod-Weinstein et al. 2020).

Partially in response to this outcry against the TMT telescope, the U.S. astronomy community's once each decade report that recommends what facilities and missions should proceed has called for the astronomy community to pivot towards "community science" (National Academies 2021). Community science encourages a larger involvement of the research community with the many other local and cultural communities participating in and affected by science projects. It also calls for greater inclusion and equity in science endeavours, in alignment with other calls for greater social justice and equity in science (O'Donnell & Scott 2021).

The science community is increasingly asking questions about which communities it belongs to and if it is a good neighbour to them. This is forcing science organisations to examine whether they have empathy for a community's problems and knowledge of its history and concerns, or if interest is only ramped up when the science institution needs something from them or encounters resistance to its plans.

**Communications, Community Engagement, Outreach, and Education: Differences that Makes a Difference**

These are actually four quite different areas, though they are often thought of as quite similar. At many research centres they are also often combined into one functional department with one set of resources to handle all of these areas. This unification is done with the notion that the four areas are closely related, can use the same people, and require the same skill sets. These assumptions are mostly incorrect; this can be best seen by an examination of each field and its audiences.

<u>Communications</u> is essential to research organisations and consists of the creation of press releases, graphics, and social media posts, and many other creative strategies to reach different audiences. Their target audience can be scientists, funding agencies, the larger organisation in which most research centres are embedded, sister institutes, the local community, the science-interested public, the government, or the public at large, to name a few.

<u>Community Engagement</u> often means encouraging a broader engagement with the science community. This can be done through expanded communications efforts including workshops, meetings, panel discussions, town halls, and strategy sessions, thus aligning scientific and non-scientific communities to support new initiatives. The beauty of the term "community engagement" is that it shifts the focus from the organisation and its output (e.g., the number of press releases) to engagement and dialog with external communities, though with a goal in mind. All community engagement efforts must first decide which community is being targeted.  Then these efforts must be sensitive to the needs of that community and to the engagement strategies that may be effective. Community engagement is usually about choosing, tailoring, and understanding the medium, message, reception, and response, and the encouragement of listening and dialogue, often to further a communication goal.

Community engagement for education is a sound and genuinely good goal but is not often done. When was the last time your favorite research organisation met with the local school district superintendent, head of local school science programs, or the indigenous leader of a nearby tribal Nation or with the First Nations community to discuss education? We strongly encourage sustained and productive interactions with the local, regional, or national cultural and educational communities to gain an understanding of their needs. In practice this is not usually done on a regular and long-term basis. Thus, community engagement efforts at research organisations have largely focused either on their own science community or on local and regional community leaders who can influence general public opinion or policy.

<u>Outreach</u> or educational outreach is often used synonymously with education when describing the efforts of research organisations. However, it is quite different as it usually refers to handling requests from outside the organisation for advice, guidance on where to look for materials, requests for tours, or requests for scientists to visit or speak. It also includes working with those outside the organisation who need guidance for their science-interested children and talking with those who want to contribute their own idea on how the universe (or their private universe) works.

Outreach covers education-related actions that require the research centre to advance outside of its walls, either physically or virtually (e.g., by video conferencing). Outreach is often justified by both its educational and public relations benefits. It is viewed as a form of social responsibility, and as such it is often viewed as an affable but not essential action. If people from the outside want to speak to someone on the inside, the public outreach department is the place to refer them to.

The priority of outreach in an organisation varies widely, from viewing it as a welcoming and much needed interface with the public to a barrier that shields and insulates the scientists from the public. A figurative example of this latter case is this: the research institute on the hill occasionally opens its formidable door and gates (and lowers its drawbridge over its moat!) and allows some lucky few from below to ascend and enter, to see the wonders that have been created. Or the institute sends its trained emissaries beyond its walls into the village below, usually bearing gifts of posters, PowerPoint slides for the Rotary Club, or props that can be used in school talks. Thankfully, most organisations realize the value of outreach and extend their efforts beyond this minimal level with great passion and enthusiasm, even if they are limited by resources

However, many organisations still conduct educational outreach with paternalistic and neo-colonial components that reflect their sense of being above or better than the community. The research institute is not in a dialogue with its local community about what the community or culture wants or needs but has decided for them on what they should want or deserve. In this case, the institute is mainly serving its own needs for connection and control at the expense of helping the community develop.

In this, there are parallels to "parachute science", where researchers from areas with more resources go to a less developed area to gather data or to use the local resources for their research. The researchers then publish the results with little or no consideration or involvement of the local population, local experts, local culture, or development of local capacity and infrastructure (DeVos 2020). The danger is that educational outreach activities may function in this same way, dropping briefly into educational settings with little awareness of educational or cultural needs, and with no sense of how to help build capacity (Heenatigala, 2022).

These outreach activities are often initiated by demands that are presented or funnelled to the outreach department by schools, clubs, or the organisation's own employees. Examples of outreach activities include a request to provide mentors or judges for a student science fair, a request for someone to participate in a career day at a school, a request for a class visit to the research facility, or an inquiry for someone to answer questions from a student or teacher. There can also be requests to provide a speaker to a group, to give a demonstration or a talk, to make a presentation to the city government, or to provide a science activity at a camp or special event. Requests are often local but can be from anywhere in the world.

These requests are taken seriously by the outreach workers, especially if they were forwarded from upper management. Responding to them takes significant time that cannot be used for other broader or more productive initiatives. Unfortunately, outreach is not usually viewed as an activity requiring special skills; the perception is that it needs only minimal knowledge coupled with abundant patience and civility. It might also be viewed as a tertiary mission of the communications department. Unfortunately, the efforts are usually viewed from the perspective of management and non-outreach professionals as "anything we do in this area will be fine and should be appreciated by those we do it for." Thus, outreach is highly reactive and relies heavily on triage. It is perceived as a favour that should be appreciated and that does not need to be evaluated or examined for its opportunity costs.

<u>Education</u> and outreach are quite different. It is tempting, and not too inaccurate to say that quality education efforts are usually everything that outreach is not. Responding to educational outreach requests is like answering your door when your next-door neighbour needs a spare rubbish bag. You should do it to show that you heard the knock and have a willingness to help. In contrast, working in education is the equivalent to mobilizing the entire block to discuss, plan, find resources, and do the things that may keep your entire neighbourhood from being bulldozed by a developer. Education relies heavily on evaluation along the way as to how you are doing, and at the end as to what level you succeeded. Education is goal-oriented (usually SMART goal-oriented) and evaluation is done against the goals. In education, the time-on-task, that is, the level of effort directly applied to something is a good indicator, as efforts are usually designed to be more prolonged and in-depth in order to be effective. Thus, educational efforts to explore a topic deeply are complementary to communications efforts designed to excite interest in a large number of people. Education programs are always designed with a sense of the opportunity costs–what cannot be done if the resources are used in particular way.

Education requires a sophisticated understanding of research and practice, and how best to interact with the complexities of the educational system (Pompea & Russo 2020, 2021a,b). Education efforts that are *ad hoc* or poorly planned are often unproductive since project success relies on the successful completion of the many parts of each project. At its core, deciding what to do in an educational context and what can be done best must rely on science education professionals with the general knowledge of how children learn and who have experience with learners of different ages. These practitioners must also have detailed knowledge of the science concepts and their accompanying misconceptions and of the pedagogical knowledge of how best a concept can be approached and explored. Thus, a strong knowledge and experience base in science and in science education are essential. As one expert in both fields described science education: "It is not rocket science, it is harder" (Nelson 2021, Pompea & Russo 2021a).

**Considerations in Science Education**
We are mainly discussing "formal" education, the education that takes place in classrooms with teachers and students. Informal, or free-choice and out of school education is also important but has been addressed in detail elsewhere (Pompea & Russo 2020). Educational efforts are always intentional; they are designed to address specific problems for well-defined educational audiences. This is necessary as schools have requirements and standards of which subjects and topics can be taught at each grade level. It makes little sense to encourage a teacher to teach a topic that shouldn't even be addressed in their class.

Education efforts are largely "operational" rather than "research and development." By operational we mean working with schools and teachers to help them in their teaching that happens every day.  R&D efforts are designed to develop or to try out a new approach and are conducted by university researchers or by expert practitioners. These R&D efforts are often funded by competitive government or foundation grants, in much the same way as science R&D gets its resources. We won't be discussing the many interesting new approaches being researched here. Instead, we focus on some basic foundational issues that form the basis of how big science organisations can be working on operational projects with local schools. Here are some basic considerations:

A. Short-term Efforts are Not Needed
The activation energy for collaboration is significant so it makes sense to partner for longer-term efforts that are more important (and more challenging) to the community. One-off activities create visibility and public relations stories, but do not satisfy the deeper educational needs of the community. Giving real value back to the community by conducting quality education projects requires long-term commitment and support. If only public relations is desired, choose easy and short projects since these project can appear successful in a press release.

B. Teachers Need Guides on the Side
The maxim "Be a guide on the side, not a sage on the stage" is applicable. Teachers need human guidance and support as much as they need teaching equipment for science. Many teachers would like to have a science resource person they can ask question to about a science topic that they don't know very well. These trust-sensitive relationships take time to develop.

C. Teachers Need and Want Additional Professional Development
Most elementary teachers get very little science training so this makes it very difficult for them to teach it effectively. Programs to give them this professional development and support afterwards are invaluable. However it takes experience to create a good programme. The good news is that the best practices for professional development in science are understood and well described in books like Loucks-Horsley et al. 2009).

D. Polite Acceptance Doesn't Mean Success
When evaluating an educational activity provided by others, many teachers will politely accept and perhaps even praise what is of only incremental value to them. This is partly because of the culture of education with its collegiality and civility, and partly because teachers are used to having their professional skills undervalued, questioned, and attacked. Because of this, further clarity is needed on whether an "outreach" effort is really of educational value to their classes or is just useful to the research organisation.

### E. Without Local Partnerships, Little Can Be Done

The close cooperation and collaboration of local stakeholders is essential to get anything done. Teachers need the permission and support of their department heads, principals, and superintendents to participate in many professional development programs. Without that support, the time for the teachers to participate may not be made available and the resources to implement additional training may be lacking, all but guaranteeing the failure of the effort. Teachers, like many other professionals, do not like programs imposed on them and especially resent programs that are poorly planned or that do not meet their needs. Detailed discussions and long-term relationships play an important role in creating understandings and in fulfilling expectations.

### F. Where Did the Project Come From?

Ask honestly whether the program originated from the community and its needs or was pitched by the science organisation to the community. There is a stark difference between finding a venue for a program that you want to do (e.g. a science lecture series by scientists) and finding out whether your needy non-scientific community really wants and needs that particular program. Projects that focus on, and mesh with the school curriculum are often the most valuable to teachers.

### G. Educational Audiences are Not All Equal

There is an essential distinction to make about choosing your educational audience. Educational programs from research institutes often go the schools where their better paid employees send their children. This is often in quite a different area than the facility. For example, the children of employees might be in schools in a suburban area nearby or in a distant borough with already excellent schools. Or perhaps the employees send their children to an already excellent private school. Conversely, if the organisation operates a research station in a rural area or on indigenous land, it is unlikely that employees send their children to the schools near the research station. When choosing an audience, ask "Where will the most value be created?" Then, aim the program at that audience. Helping the wrong school is gilding the lily. There is no need for excess, when many are lacking the most basic.

### H. Some Educational Projects are Hard

An educational program, if well designed, will have the most value by going to a place where it is needed desperately. However, going to such a place is not easy. For example, many city-centre schools and schools in core neighbourhoods have started to resemble prisons in the United States, and access by outside visitors is restricted. These programs may be more frustrating and take more effort. Some programs are designed to pluck the easy to reach, low-hanging fruit. But don't confuse that kind of success with that from programs that are aimed at larger and harder to reach goals, and which may have great rewards if they succeed.

I. Projects Rely on Co-Creation to Meet Participant Needs
Developing authentic partnerships for education is much harder than creating your own outreach events or educational programs independently of your audience or users. For example, a research institute's program that creates teacher training workshops without direct involvement of teachers and the school district is sending this message:

> "We think there is a need for what we want to do. We have decided by ourselves to proceed to do this. We have designed something by ourselves, since this was easier. We think highly of our own approach and choices, and we hope you like it, too. Who wants to come and get this training? When you finish, we hope it was helpful, but we would rather not know too much. We will do it again sometime, as we feel that is was useful to you. Thank you for telling us that it was useful." (See previous item D for the translation of polite statements by teachers.)

J. Elementary Schools are the Most Needy
Teaching at the elementary school level is the most challenging type of teaching (Pompea & Russo 2021a) and incorporating science into these classrooms has always been particularly difficult. Spend time in these classrooms if you are sceptical. Elementary teachers often have large classes and many different kinds of learners who are at many different levels. The teacher is responsible for many areas of instruction and is under extreme pressure in many cases. Encouraging robust science education programs in these classrooms is critically important (National Academy of Sciences 2022).

**The Situation of Science Education Today**
Research labs often interact most strongly with upper-middle class suburban and private schools and draw their conclusions on the educational system from those experiences. The needs of those schools and their students may not be typical. Here are some common concerns from more typical public schools worldwide:

- Public schools have inadequate technology, a lack of books, and poor or no science lab equipment.

- There is a shortage of trained teachers in the science disciplines at all grade levels. Many science teachers are teaching outside their field of expertise.

- There is a strong disconnect between science classrooms and how science is done, with many science teachers having little practical experience in doing science. (Pompea & Russo 2021a)

- Elementary and primary grade teachers have little training in science content or in science teaching pedagogy and may not teach very much about science.

- Teaching has not attained the full status of a profession or guild (Tobias and Baffert 2010). Disrespect for teachers and their lack of autonomy seems to be getting worse each year.

In the United States there are additional problems. School segregation and unequal school financing is still ubiquitous. The emergence of charter schools and the expansion of school voucher systems have undermined a once strong public education system. Government leaders such as former Secretary of Education Betsy DeVos and her predecessors have also contributed to the undermining of public schools by promoting the contrived virtues of private schools over public schools (Ravitch 2014, 2016, 2017). In the United States, the number of teaching certificates awarded for secondary school teaching in science, technology, engineering, and math is dropping precipitously (Physics Today 2022).

Another large problem in the U.S., but that also occurs worldwide, is the increased attack on science and science education, often by the same groups attacking the value and autonomy of public education. The "Scopes Monkey Trial" of 1925 convicted a science teacher of violating a Tennessee law prohibiting the teaching of evolution. Although this law was taken down in recent years, there is very little teaching of biological evolution in many states. Organisations like the U.S. National Academy of Sciences have worked hard to counteract such teaching prohibitions (National Academy of Sciences 1998) but many issues remain.

Laws are now being passed to hinder teaching about the ancient Earth, the Big Bang, and climate science. The argument is that these are inappropriate "controversial" topics and should not be taught in schools. There are numerous proposals to require schools to teach "creation science" or "intelligent design" in science classrooms. Most U.S. textbook publishers have significantly altered school science textbooks to make them more acceptable to anti-science textbook committees. This situation will not solve itself.

The current attacks on science and the science of global climate change have tremendous repercussions for our planet. The next generation of children must have the opportunity to develop an understanding and appreciation of how science works and the value of scientific data and evidence in order to address this important issue.

**The Business Model Perspective**
A research institute can help itself short-term (i.e., achieve a good return on investment) by vigorously communicating its successes: its science discoveries, service to its users, and the relevance of its research to society (Entradas et al. 2020).  Similarly, from a short-term business perspective, research institutes can be effective in generating positive community responses from short-term, *ad hoc*, self-serving, or unwanted educational efforts. Is this the right way to evaluate a research institute's efforts? That value can be obtained by these educational efforts perhaps illustrates that the community is grateful for the smallest efforts made, especially when none have been made before.

However, the gains obtained through self-serving approaches rarely help develop the deeper community trust or critical mass of supporters in big science projects (and their sponsoring institutions) that is essential today for many ambitious big science projects. The process of building an iterative and equitable relationship with a community is much more difficult. (Kahanamoku et al. 2020).

**Adding an Ethical Perspective**
We feel that it is time for research institutes to move beyond basing their educational actions on the direct return for the organisation and its stakeholders (Entradas et al 2020). This type of justification can be convincing, but it is also rather narrow. The case for promoting diversity in organisations has often been justified in this same narrow way, namely that diversity improves the productivity of the organisation. While diversity does improve the productivity, adding ethical considerations can strengthen the argument for encouraging a diverse work force (Haacker et al. 2022).

Ethical considerations are a key part of many decisions and adding it to strategic planning for education broadens the perspective. Ethical considerations allow us to examine the moral questions and dilemmas associated with the choices of action or non-action that can be made by research organisations as they begin to notice the local education arena and its needs.  Hard questions need to be addressed:

> "Do individuals and their organisations have an ethical obligation to address and repair the problems of the world, or at least the ones closely related to their profession, culture of science, and support for science?"
>
> "If the local science education situation is seriously unwell, is there a duty to help or rescue it, even if the rescue effort is hard?"

Can, science organisations behave like the ploughman in Auden's 1940 poem "Musée des Beaux Arts", who chose not to notice the fall of Icarus? These questions frame the ethical considerations that need to be considered in addition to considering the direct and short-term value that comes back to the organisation from its education efforts.

**Educational Stewardship: Preserving the Educational Landscape**
A second perspective that can be added to the ethical perspective is that of "stewardship" of science education and of the public school system. A culture of stewardship and partnership can frame how research organisations support and improve the landscape of science education.

Just as environmental stewardship is a key element in the maintenance of fragile places, the adoption by research institutes of an educational stewardship approach will help them work with others to maintain the public schools and to reinforce their important role in our society. A stewardship approach also can help support the culture of scientific research and discovery so essential to the future of research. With some reflection, we should realize that there are strong long-term connections between science teaching and science research, that both are vulnerable, and that by working together they can both flourish.

The protection and caring for land is one of the original meanings of stewardship. This caring perspective includes all elements of the present land, and the cleaning up and repairing of damage from earlier times. In true stewardship, this responsibility exists even if the present stewards didn't cause the damage or realize its extent. Stewardship implies a responsibility to preserve for future generations and includes making decisions that avoid short or long-term harm. The research literature on land-use planning and stewardship is instructive in its descriptions of how land can be protected and nurtured in the face of diverse threats (Wallace et al. 2005).

For educational stewardship, the same general principles of land stewardship apply. Educational stewardship can preserve the landscape of the public education system and its main inhabitants: the teacher, students, and administrators. It involves maintaining and improving the parts of the educational landscape that have not been taken care of previously. This culture of stewardship is not about replacing public education with private versions. As in land stewardship, stewardship demands a commitment to preserving what is there, even if has been been degraded under the guise of educational reform or improvement.

Research labs and large user facilities are often based in rural areas or on indigenous land. This is particularly true for astronomical observatories like the Square Kilometer Array in South Africa and Australia and for high energy physics facilities, like the Pierre Auger Observatory in Argentina. Rural and indigenous communities often struggle with poverty, youth unemployment, low teacher retention, poor internet service, and overwhelmed health facilities. The Covid-19 pandemic has accentuated many of these problems even further. In particular it has highlighted the importance of computers and the internet in providing equitable educational access.

At first glance, research institutes seem ill-prepared to become stewards of either the public school system or the science education ecosystem since they have few education professionals on their staff. However, "Big Science" projects and the facilities that support them are actually quite rich in resources that are valuable for educational stewardship and even for the broader stewardship of our planet. In particular, the interest and time of their employees is a valuable resource.

Science organisations have a large number of highly educated scientists, engineers, and facility users as well as a diverse and often international work force. They also have a strong supporting infrastructure of management, facilities, purchasing, and accounting that can support organisational missions. Finally, their core values are highly congruent with the core values in education. These attributes can be used to great advantage in education and community engagement efforts. The longevity of research facilities also can contribute to developing strong community roots and allegiances, which are essential to stewardship.

**What Research and Development Organisations Can Do: A 3 + 5 Proposal**
Our starting (and ending) point is that research institute resources should be used in the schools and other situations where the need of learners are the greatest. In our experience this means the elementary or primary public schools, i.e., the schools that serve students who are 11 years old or younger. Educationally vulnerable students are often found in the postal code areas that are less affluent. Students at private schools or a public school situated in an upper middle class neighbourhood are usually much better served. Public schools that serve inner city, rural, and indigenous residents usually have the most needs and a clear case for equity and fairness can be made. Where there is a dire situation for the students, that is where the help is needed, and also where the help can do the most good, in the long-run.

We propose a 3 + 5 programme to describe a moderate level of educational support that can be productive. If research organisations and the companies that provide support to them could commit to 3% of their budget to education and 5% of any employee's time to work on the organisation's educational programs, this would be a step in the right direction. This is the essence of the program: a minimum of 3% of the budget and 5% of the time for some employees.

In this plan for action, 3% of the regular or core budget would be solely for education staff and the resources they directly need. Not included in this 3% would be money people or time for organising, creating, and supporting booths, meetings and conferences; town halls; press releases; social media production; graphics production; web site improvements; working with film crews; arranging interviews with media; rotary club talks; or for answering communications inquiries. It would be used to hire or support education professionals who can organise and direct the broader range of truly educational activities for the staff who are interested. The money can also be used to provide the resources needed for these efforts. Grant money on top of the 3% can be used to organise special projects or educational experiments.

The second number of 5% represents the amount of regular time allocated for staff who are motivated to participate in these organised and productive educational activities. These are the activities that were created in a dialogue with the local stakeholders. If even 5-10% of the organisation's employees participate in this program, 5% of their time is a significant effort that can supplement the efforts of the educational specialist.  An employee who has 5% of their time available for education work means that they can spend one full day a month preparing and actually working with schools or teachers. This represents a great improvement over having one or a few education professionals in a large organisation trying to work with schools or students.

The staff volunteers can start that day by not coming to work and spending time in a school, learning their needs. Note that this 5% of time is not allocated for working at their own children's school (unless it is a highly needy one), or in some out of school activity that they feel strongly about.

We don't discourage these volunteer activities, but this time is meant to be in direct service of their parent organisation in a well-planned program. This time is for employees to participate in some solid activities or projects organised and supervised by their own education professional(s). This might be a tutoring program, or it might be to support a professional education and training program for teachers, for example.

Paid employee time is a powerful resource and represents a deep organisational commitment that can augment the expertise of the organisation's education professionals. Allocating and freeing the time for employees who wish to participate to do educational community service is critical, and represents a serious commitment. Encouraging employees to volunteer time in the evenings and on weekends for educational causes is not ineffective, but it gives the message that these efforts are not a priority.

If these 3 + 5 programs are implemented, they need to run long-term, and not be subject to budget cuts or attempts to subvert it through merging of budgets and other common ways that shift operational priorities.

**Moving from Local Support to National Support**
The allocation of resources to contribute locally to schools and the local educational ecosystem is a most important part of the picture. Many organisations have created successful local educational partnerships and have then gone beyond these local contributions in order to contribute on a larger scale. The national and international character of big science projects often pushes a project to contribute beyond the local level as the organisation evolve educationally and partner more widely (Pompea et al 2015).

There are many examples from our own experiences of how research organisations may contribute to larger regional, national, or international efforts to promote science education. We will use a few from our field of physics and astronomy that may serve as models for other fields of science. We especially encourage the development of programs that encourage science interest and develop science skills in younger and pre-secondary school children.

For example, *Universe Awareness* (UNAWE) is an international program for elementary school students (Miley et al. 2019) that trains teachers and develops inexpensive innovative materials for classrooms. The *Hands-On Optics* program was a United States program funded by the National Science Foundation for 11-13 year old school students. It developed innovative curricular materials and kits and trained museum educators and teachers across the country (Pompea et al. 2005). The *Galileoscope* program (Pompea et al. 2010) was an international program to develop and teach with small, high-quality telescope kits. It was also used in Arizona as a state-wide program for 10 year old students who built their own telescopes and used them at star parties (Pompea et al. 2013, 2015). Some projects, such as the *Colors of Nature* project focus specifically on new approaches that encourage girls in science activities (Tzou et al 2014; Conner et al. 2017).

Other innovative projects with global reach that were developed at research organisations include the *International Year of Astronomy 2009* (Russo et al. 2009) and the *International Year of Light 2015* (Dudley & González 2015; Curticapean et. al. 2016). Additionally, there are many educational initiatives such as the *IAU100* and *Pale Blue Dot* led by professional societies such as the International Astronomical Union (van Dishoeck & Elmegreen 2018; van Dishoeck 2019). A number of highly successful programs for older students have emphasized research using observations and data from research telescopes on the ground and in space (Buxner 2010; Doran et al 2012; Fitzgerald et al. 2014; Rebull et al. 2018a,b). Many other exemplary programs are described in Pompea & Russo (2020). These programs have been highly effective either to encourage science interest or to suggest new approaches that can be incorporated into current education programs.

**Conclusion and Summary**

We have described the resources big science has to offer to its local community and to the larger science education ecosystem. These resources are valuable and need to be allocated and committed in an intentional way for effective results. The efforts need to be given to those who can benefit the most from them, namely public elementary school students and their teachers.  Educational projects will only succeed if they are based on the best practices elucidated by science education research. A key best practice is to form strong, authentic partnerships and to co-create programs with educational stakeholders and organisations. Stronger educational engagement with the local, regional, national, and international education community is not only possible, but highly desirable from an ethical and stewardship perspective.

We've suggested a sensible allocation of money and time to implement strong local educational partnerships in our 3 + 5 approach. For some research organisation this appears to be a large stretch to implement such a program. These organisations feel vulnerable working in this new direction and worry about failing. This fear should not be an impediment to getting started.  Perhaps they should take heart from the words of a doctor who worked under very difficult circumstances. Paraphrasing the Talmud, he said: "The day is short. The task is difficult. It is not possible to complete it. But we are forbidden not to try" (White 2013).

We encourage big science organisations to try, to commit themselves to the difficult task of supporting science education. Although there are easier ways for an organisation to look good to their community in the short-term, the long-term benefits are powerful and important to the science organisations, to their people, and to the new communities they now will serve.

**Acknowledgements**
The work of SMP is supported by NSF's NOIRLab, which is managed by the Association of Universities for Research in Astronomy (AURA) under a cooperative agreement with the National Science Foundation.

The opinions expressed by the authors are theirs alone and not that of NSF's NOIRLab, AURA, Inc., the National Science Foundation, the University of Arizona, or Leiden University.

National Academy of Sciences (1998), *Teaching About Evolution and the Nature of Science*, Washington, DC: National Academies Press. Available at: http://nap.edu/catalog/5787.html.

National Academy of Sciences 2017, *Communicating Science Effectively: A Research Agenda* (2017) Washington, DC: National Academies Press.

National Academy of Sciences 2018, *The Science of Science Communication III: Inspiring Novel Collaborations and Building Capacity: Proceedings of a Colloquium,* Washington, DC: National Academies Press.

National Academies of Sciences, Engineering, and Medicine, 2021, *Pathways to Discovery in Astronomy and Astrophysics for the 2020s*.

National Academies of Sciences, Engineering, and Medicine, 2022, *Science and Engineering in Preschool Through Elementary Grades: The Brilliance of Children and the Strengths of Educators*, Washington, DC: The National Academies Press. https://doi.org/10.17226/26215.

Nelson, G. "Former Astronaut George 'Pinky' Nelson Champions Science Literacy," e-interview, Education World (accessed 5 August 2021).

NSF Award 1447720, 2022, University of California Irvine Michael Carey P.I. Carey Retrieved March 10 2022 at https://www.nsf.gov/awardsearch/showAward?AWD_ID=1447720

O'Donnell, C. and Scott, K.A., "Culturally responsive astronomy education: using a critical lens to promote equity and social justice." Retrieved from https://caodonnell.github.io 8 April 2022.

*Physics Today* "The US is in dire need of STEM teachers" 2022, **75**, 3, 25 ; doi: 10.1063/PT.3.4959

Pompea, S. M., Johnson, A., Arthurs, E., & Walker, C. E. 2005. "Hands-On Optics: an educational initiative for exploring light and color in after-school programs, museums, and hands-on science centers", in Proc. Ninth International Topical Meeting on Education and Training in Optics and Photonics, Marseille, France.

Pompea S. M., Schweitzer, A., Deustua, S., Isbell, D., Fienberg, R. T., Arion, D. N., Walker, C.E., Gay, P.L., Smith, D., Pantoja, C.A., Watzke, M., and Arcand, K., 2010. "International Year of Astronomy 2009 Cornerstone Projects: What's Available for You", *Science Education and Outreach:*